\documentclass[a4paper,10pt]{article}

\pdfoutput=1

\usepackage{cite,amsmath,amsfonts,amsthm,fullpage}
\usepackage{youngtab}

\usepackage{graphics}
\usepackage{miniplot}
\usepackage{subfigure}

\newcommand{\Pa}{\mathop\mathrm{P}\nolimits}

\theoremstyle{plain}

\newtheorem{Lemma}{Lemma}
\newtheorem{Proposition}{Proposition}
\newtheorem{Corollary}{Corollary}
\newtheorem{Remark}{Remark}
\newtheorem{Example}{Example}

\theoremstyle{remark}

\def\Tr{\mathrm {Tr}}

\def\det{\mathrm {det}}

\def\bp{\begin{Proposition}}
\def\ep{\end{Proposition}}
\def\bc{\begin{Corollary}}
\def\ec{\end{Corollary}}
\def\bl{\begin{Lemma}}
\def\el{\end{Lemma}}
\def\be{\begin{equation}}
\def\ee{\end{equation}}
\def\br{\begin{Remark}\rm\small}
\def\er{\end{Remark}}
\def\brs{\begin{remarks}.\\ \rm\
\begin{enumerate}}
\def\ers{\end{enumerate}\end{remarks}}
\def\bea{\begin{eqnarray}}
\def\eea{\end{eqnarray}}

\def\bx{\begin{Example}\rm\small}
\def\ex{\end{Example}}

\def\Tr{\mathrm {Tr}}
\def\tr{\mathrm {tr}}
\def\det{\mathrm {det}}

\def\&{&{\hskip -20pt}}

\newcount\YDcount\YDcount=0
\def\YDsize{10pt}

\def\YD#1{%
\ifnum#1=0
 \ifnum\YDcount=0 \ifx\varnothing\undefined\emptyset\else\varnothing\fi
 \else\vskip1.4pt\egroup\YDcount=0\fi
\else
 \ifnum\YDcount=0 \YDcount=1\vcenter\bgroup\vskip1pt
 \else\nointerlineskip\fi
 \vbox{\hrule\hbox{\vrule height\YDsize
 \loop\hskip\YDsize\vrule\ifnum\YDcount<#1\advance\YDcount1\repeat}\hrule
 \kern-0.4pt}\expandafter\YD
\fi}

\usepackage[usenames,dvipsnames]{color}
\usepackage{ulem}

\def\pb{\mathbf{p}}
\def\tb{\mathbf{t}}

\begin{document}

\author{ C. Z. Li \thanks{College of Mathematics and Systems Science, Shandong University of Science and Technology, Qingdao,
266590, China\\ {\bf email:} lichuanzhong@sdust.edu.cn}\\
A. Mironov \thanks{Lebedev Physics Institute, Moscow 119991, Russia\\ NRC ``Kurchatov Institute", 123182, Moscow, Russia\\ Institute for Information Transmission Problems, Moscow 127994, Russia\\ {\bf email:} mironov@lpi.ru, mironov@itep.ru}\\
A. Yu. Orlov\thanks{Institute of Oceanology,
Moscow, 117997, Russia\\
NRC Kurchatov Institute, Moscow, 123182, Russia\\
Institute for Information Transmission Problems,
Moscow, 127051 Russia\\
National Research University Higher School of Economics,
Moscow, 101000 Russia\\
{\bf email:} orlovs@ocean.ru}
}

\title{{\bf Tau functions of the UC hierarchy as partition
functions of matrix models }}

\maketitle

\vspace{-8.5cm}

\begin{center}
  \hfill FIAN/TD-12/25\\
  \hfill ITEP/TH-23/25\\
   \hfill IITP/TH-20/25
\end{center}

\vspace{6cm}

\begin{abstract}
We present a family of matrix models such that their partition functions are tau functions of the universal character (UC) hierarchy. This develops one of the topics of our previous paper arXiv:2410.14823. We found new matrix models associated with the product of two spheres with embedded graphs via a gluing matrix. We also generalize these studies to multi-matrix models case, which corresponds to the multi-component UC hierarchy.
\end{abstract}

\bigskip
\medskip

\tableofcontents


\section{Introduction \label{Introduction}}

More than half a century passed from the basic discoveries concerning
wonderful properties of integrability which were resumed in the famous book
\cite{edNovikov} edited by Sergei Novikov (see also the book \cite{BC} with the Russian translation edited by Sergei Novikov). Since then, the topic was developed in many directions.
One of these directions was an application to theory of random matrices \cite{Wigner,Dyson,Mehta}, which
possesses a lot of links with various problems of mathematics, physics, statistics
and transmission of information. Here we describe a relation between one of the modification of integrable hierarchies called UC hierarchy \cite{Tsuda} and a few new families of exactly solvable matrix models. ``Exactly solvable'' means that the related integrals can be written as a tau function. UC tau functions introduced in \cite{Tsuda} can be written in the form of special series of the
so-called universal characters (UC) associated with pairs of partitions.

We begin with a remark about our notation: throughout the paper, we use the
power sum variables $\pb=(p_1,p_2,\dots)$ common in the theory of symmetric functions instead of
the time variables $\tb=(t_1,t_2,\dots)$ common in the theory of integrable systems, and also call the variables
$p_1,p_2,\dots$ times.
The relation between the two sets of variables is $p_m=mt_m$. We also deal with the Schur functions, the universal characters and tau functions as functions of these power sum variables.

Matrix models can be represented as the mathematical expectation of an exponential:
$$
\langle
\exp \sum_m \frac{p_m}{m}\tr \left(X^m  \right)
\rangle_\Omega ,
$$
where $X$ is a matrix belonging to some ensemble of random matrices $\Omega$
\footnote{The detailed description of various ensembles of random matrices can be found in \cite{Mehta}}, and
$p_m$ are parameters.
More general matrix models are constructed as averages over multimatrix ensembles (e.g., see \cite{Akeman}), and it is not exponentials that are averaged, but more complex expressions (e.g., tau functions can be averaged, see Appendix in \cite{HO-2005}, and also \cite{NO}).

The parameters of a matrix model can be either sets of numbers, which we call coupling constants, or sets of constant matrices, which we call source matrices. A well-known example of such a model, depending on the source matrix, is the Kontsevich integral \cite{Kontsevich}.

Thus, the new families of matrix models that we introduce in this paper are tau functions, which are defined either as functions of coupling constants (then the coupling constants play the role of times), or as functions of source matrices (then one obtains tau functions in Miwa variables). We use the expansion of matrix integrals into the above-mentioned universal characters. These characters can respectively be functions of times, or can be functions of source matrices, which are two different cases that our further consideration is based on.

If the set of power sum variables $\pb=(p_1,p_2,\dots)$ is parametries by a matrix ${\cal V}$
\be\label{miwa}
p_m=p_m({\cal V}):=\tr \left({\cal V}^m\right),\quad m=1,2,\dots
\ee
 then, the eigenvalues of ${\cal V}$
we will call Miwa variables, such terminology is of use in works devoted to
classical integrability. When we say a tau function is written in Miwa variables we
mean that $\tau(\pb)=\tau(\pb({\cal V}))=:\tau({\cal V})$, we hope it will not produce a misunderstanding, we use capital letters for matrices and bold letters for the collection of power sum variables.

\paragraph{Schur functions and universal characters.} It is well-known that
the character of the polynomial irreducible representation of linear group labeled by
a partition $\lambda=(\lambda_1,\dots,\lambda_{\ell})$ is given by the formula
$$
s_\lambda(X)=
\frac{\det\left[x_i^{N+\lambda_i-i}  \right]_{i,j\le N}}
{\det\left[x_i^{N-i}  \right]_{i,j\le N}},\quad X\in \mathbb{GL}_N(\mathbb{C}),
$$
where $x_i$ are eigenvalues of $X$. The polynomial $s_\lambda$ called Schur function is a symmetric
function of $x_i$. If one introduces an infinite set of power sum variables
$\pb=(p_1,p_2,\dots)$ \cite{Mac}
\be\label{powerSums}
p_n =\tr\left(X^n \right) ,\quad \deg p_n=n,\quad n=1,2,\dots
\ee
then the Schur function can be re-written as a quasi-homogeneous polynomial ($\deg s_\lambda=|\lambda|:=\sum_i \lambda_i$) in the variables $\pb$
in the following way:
\be\label{Schur-in-p}
s_{\lambda}({\bf p})=\det\left(s_{(\lambda_{i} -i+j)}({\bf p})
\right)_{1\le i,j \le N'},\quad N'\ge \ell(\lambda),
\ee
which is independent of $N'$ if we put $s_{(k)}=0$ for $k<0$. Hereafter, we denote through $\ell(\lambda)$ the length of the partition $\lambda$, see Appendix \ref{Partitions}.

The relation (\ref{Schur-in-p}) can be treated as a definition of the Schur function
in the variables $\pb$ without any relation with the linear group and the variables $x_1,\dots,x_N$
if one defines the set of $\{s_{(n)},\,n=1,2,\dots\}$ entering (\ref{Schur-in-p}) though the formula
\be
e^{\sum_{n>0}\frac 1n p_n x^n}=\sum_{n\ge 0} s_{(n)}(\pb) x^n.
\ee
We use the same notation for $s_\lambda$ for the Schur functions as functions of different variables: both $\pb$ and $x_1,\dots,x_N$, hopefully this will not cause confusion.

Here, it is suitable to recall the formula which defines
the skew Schur function \cite{Mac}:
\be\label{skew-Schur-in-p}
s_{\lambda/\mu}({\bf p})=\det\left(s_{(\lambda_{i} -i-\mu_j+j)}({\bf p})
\right)_{1\le i,j \le N'},\quad N'\ge \ell(\lambda),\ell(\mu).
\ee
In particular, $s_{\lambda/0}=s_\lambda$. It is supposed that the Young diagram of the partition
$\mu$ is a subdiagram of the Young diagram of partition $\lambda$, details see in \cite{Mac}.

Now, consider two sets of variables $\pb$ and $\pb^*$ where $\deg p_n=n$ and $\deg p^*_n=-n$.
The universal character is labeled by a pair of partitions $\lambda=(\lambda_1,\dots,\lambda_l)$ and $\mu=(\mu_1,\dots,\mu_{l'})$ and can be defined as the following quasi-homogeneous polynomial
\be
s_{[\lambda,\mu]}({\bf p},{\bf p}^*)=\det\left(s_{\mu_{l'-i+1} +i-j}({\bf p})  ,\,1\le i\le l'
\atop s_{\lambda_{i-l'}+i-j}({\bf p}^*),\, l'< i\le l'+l \right)_{1\le i,j \le l+l'},\quad
\deg s_{[\mu,\lambda]}=|\mu|-|\lambda|.
\ee
When
\be\label{p(X)}
p_n=p_n(X)=\tr \left(X^n\right),\quad p^*_n=\tr\left( Y^n\right),\quad n=1,2,\dots
\ee
one writes
\be\label{UC(X,Y)}
s_{[\mu,\lambda]}(X,Y):=s_{[\mu,\lambda]}({\bf p}(X),{\bf p}^*(Y)).
\ee
The notion of the universal character originally appeared as a generalization of the character of the rational representation
of the linear group, see \cite{Koike} and \cite{Tsuda}. In this case, $X$ and $Y$ are related:
$Y=X^{-1}\in \mathbb{GL}_N$.

In \cite{Tsuda}, there were written down equivalent definitions of universal characters as
\be\label{def2}
s_{[\mu,\lambda]}(\pb,\pb^*)=\sum_{\nu} (-1)^{|\nu|}s_{\mu/\nu}(\pb)s_{\lambda/\nu^t}(\pb^*),
\ee
where $\nu^t$ denotes the partition conjugated to $\nu$ (see \cite{Mac}) and where $s_{\mu/\nu}$ are the skew Schur functions \cite{Mac},
and also as
\be\label{mating0+}
s_{[\mu,\lambda]}(\pb,\pb^*)=
s_\mu(\pb-\partial^*)\cdot s_\lambda(\pb^*),\quad
\partial^*:=(\partial^*_1,2\partial^*_2,\dots),\quad \partial^*_m=m\frac{\partial}{\partial p^*_m}.
\ee

Formulas (\ref{mating0+}) and (\ref{def2}) are equivalent since
 $s_\lambda(\tilde\pb-\partial)=\sum_\nu (-1)^{|\nu|}s_{\lambda/\nu}(\tilde\pb)s_\nu(\partial),$
and $s_\nu(\partial)\cdot s_\mu(\pb)=s_{\mu/\nu}(\pb)$
(see \cite{Mac}).

Formula (\ref{mating0+}) provides a procedure to build the universal character from a pair of
Schur functions, we call such a procedure of getting the universal character {\it mating}
of a pair of the Schur functions $[s_\mu \ast s_\lambda]=s_{[\mu,\lambda]}$.

\paragraph{Matings of pairs of the Schur functions}

We define three ways of mating $[f\ast g]$ of a pair of functions $f$ and $g$, denoted by $[f\ast g]_1$,  $[f\ast g]_2$ and $[f\ast g]_3$,
where the first two of them are related to the functions of power sum variables $f=f(\pb),\, g=g(\pb^*) $, and the last one mates functions of matrices $f=f(X),\,g=g(Y)$:

\begin{itemize}
\item[(i)] $[f\ast g]_1$:
\be\label{m0}
e^{-\sum_{n>0} n \partial_n\partial^*_n}f(\pb)g(\pb^*)=[f(\pb)\ast g(\pb^*)]_1:=f(\pb-\partial^*)\cdot g(\pb^*)\equiv
[g(\pb^*)\ast  f(\pb) ]_1=
g(\pb^*-\partial)\cdot f(\pb),
\ee
where
\be
\partial:=(\partial_1,\partial_2,\dots),\quad \tilde\partial:=
(\tilde\partial_1,\tilde\partial_2,\dots),\quad \partial_m=m\frac{\partial}{\partial p_m}
,\quad \tilde\partial_m=m\frac{\partial}{\partial \tilde p_m}.
\ee

In particular $[s_\mu(\pb)\ast s_\lambda(\pb^*)]_1=[ s_\lambda(\pb^*)\ast s_\mu(\pb)]_1 =s_{[\lambda,\mu]}(\pb,\pb^*)$.

\item[(ii)] $[f\ast g]_2$:
\be\label{m1}
[f(\pb)\ast g(\pb^*)]_2:=
\int_{\mathbb{U}_N}
f\left(\pb+\pb(U)\right)g\left(\pb^*-\pb(U^\dag)\right)d_*U
\ee
where $d_*U$ denotes the Haar measure on the unitary group $\mathbb{U}_N$ and
where the notation $\pb(U)$ was defined in (\ref{p(X)}).
One can prove that $[f(\pb)\ast g(\pb^*)]_2= [ g(\pb^*)\ast f(\pb)]_2$.
In particular,
\be\label{mating1+}
\left[s_\mu(\pb) \ast s_\lambda(\pb^*)  \right]_2:=
\int_{\mathbb{U}_N}
s_\mu\left(\pb+\pb(U)\right)s_\lambda\left(\pb^*-\pb(U^\dag)\right)d_*U=
\ee
$$
=\int_{\mathbb{U}_N}\sum_{\nu}s_{\mu/\nu}(\pb)s_\nu(U)
\sum_{\alpha}s_{\lambda/\alpha}(\pb^*)s_\alpha\left(-\pb(U^\dag)\right)d_*U=
$$
$$
=\sum_{\nu,\alpha}s_{\mu/\nu}(\pb)s_{\lambda/\alpha}(\pb^*)(-1)^{|\alpha|}
\int_{\mathbb{U}_N}s_\nu(U)s_{\alpha^t}(U)d_*U
$$
\be\label{answer[ss]_2}
=\sum_{\nu\atop \ell(\nu),\ell(\nu^t) \le N}(-1)^{|\nu|}s_{\mu/\nu}(\pb)s_{\lambda/\nu^t}(\pb^*),
\ee
where we indicate the summation range explicitly (which follows from the fact that $s_\nu(U)=0$ in case
$U\in {\rm Mat}_{N\times N}$ and
$\ell(\nu)>N$) in order to relate this formula to (\ref{def2}). Then, one observes
that in general the right hand side is not equal to the universal character due to the restriction
$\ell(\nu)\le N$ in the summation range: in order to have the equality (\ref{def2}), one
needs the sum over {\it all} partitions $\nu$ (though it is naturally restricted
by the relations of type $s_{\lambda/\nu}=0$ if the partition $\nu$ is not a sub-partition
of $\lambda$), and, in the case of small enough $N$, there is not a sufficient number of terms in
(\ref{answer[ss]_2}) to get the whole sum in (\ref{def2}).

We have the following sufficient (though not necessary) condition to obtain the universal character
from the mating (\ref{mating1+}) of two Schur functions:
\bl\label{lemma1}
When the both $\ell(\lambda)$ and $\ell(\mu)$ do not exceed $N$, one gets
\be\label{cond-for1}
\left[s_\lambda(\pb) \ast s_\mu(\pb^*)  \right]_2=s_{[\mu,\lambda]}(\pb,\pb^*).
\ee
\el
The proof follows from the fact that
the intersection of partitions $\lambda$ and $\mu^t$ is completely situated
within the square $N\times N$, which is the summation range of the partition $\nu$ in (\ref{def2}).

Let us note that the requirement that the first parts of the partitions, $\lambda_1$ and $\mu_1$ do not exceed $N$ is enough to guarantee
the relation (\ref{cond-for1}).

One can also see that $[s_\mu(\pb) \ast s_\lambda(\pb^*)]_2=[s_\lambda(\pb)\ast s_\mu(\pb^*) ]_2$.

\item[(iii)]
Two previous matings were defined for functions of the power sum
variables. Now, we introduce the mating of functions of matrix argument, $\left[f(A_1)\ast g(A_2)\right]_3$:
\be
\left[ f(A_1) \ast g(A_2)\right]_3=\int_{{\mathbb U}_n\times {\mathbb U}_n } dU_1 dU_2 \prod_{a=1,2}\det\left(I_{n}\otimes I_{n}-U_1\otimes U_2\right)
 f\left(A_1(X_1,U^\dag_1)\right)g\left(A_2(X_2,U^\dag_2)\right),
\ee
where the both matrices $A_1$ and $A_2$ have the following block structure
 ($a=1,2$):
\be\label{block}
\left(A_a\right)_{i,j}=\left(A_a\right)_{i,j}(X_a,U^\dag_a)=\begin{cases}
                              (U^\dag_a)_{i,j}\quad i,j \le n, \\
                            (X_a)_{i,j}\quad n<i,j\le N-n, \\
                            0\quad {\rm otherwise}.
                             \end{cases}
\ee

Note that, in this case,
\be\label{skew}
s_{\lambda}(A_a)=\sum_{\nu} s_{\lambda/\nu}(X_a)s_\nu(U_a).
\ee
Then
\bl\label{l1}
\be
\int_{{\mathbb U}_n\times {\mathbb U}_n } dU_1 dU_2
s_\nu(U_1)s_{\nu^t}(U_2) s_\lambda(A_1)s_\mu(A_2)  = s_{\lambda/\nu}(X_1)s_{\mu/{\nu}^t}(X_2).
\ee
\el

One also has
\bl
\label{l2}
Let $U_{i}\in\mathbb{U}_{n}$, $i=1,2$.
\be
{\cal G}(U_1,U_2):=\det\left(I_{n}\otimes I_{n}-U_1\otimes U_2\right)=
e^{-\sum_{k>0}\frac1k\tr U_1^k\tr U_2^k}
=\sum_{\nu\atop \ell(\nu),\ell(\nu^t)\le n} (-1)^{|\nu|} s_\nu(U_1)s_{\nu^t}(U_2).
\ee
\el

Let us introduce the following notation:
\be\label{UC^[n]}
s^{[n]}_{[\lambda,\mu]}(X_1,X_2):=\sum_{\nu\atop \ell(\nu),\ell(\nu^t)\le n} (-1)^{|\nu|}
s_{\lambda/\nu}(X_1)s_{\mu/{\nu}^t}(X_2)
\ee

From these Lemmas and from (\ref{def2}), one obtains

\bl\label{mate-2-UC}
In the case of $\ell(\lambda),\ell(\mu)\le n$, one gets
\be\label{cond-for2}
\left[s_\lambda(A_1) \ast s_\mu(A_2)  \right]_3\equiv s^{[n]}_{[\lambda,\mu]}(X_1,X_2)=s_{[\mu,\lambda]}(X_1,X_2).
\ee
\el

\end{itemize}

\paragraph{KP and UC tau functions.}

A remarkable observation by Sato is that the Schur function is a solution
to the famous KP hierarchy \cite{Sato1}, which can be written in a bilinear form (Hirota equations).
In a similar way, the universal
character solves another integrable hierarchy introduced by Tsuda \cite{Tsuda}
and called universal character hierarchy, or just UC hierarchy.
This topic was developed in a number of works \cite{LMO1,TsudaPainleve,TsudaDiff}.

We recall that the Sato formula for the KP tau function depending on the set of times
$\pb=(p_1,p_2,\dots)$ is as follows:
\be
\tau^{KP}(\pb)=\sum_\lambda \pi_\lambda s_\lambda(\pb).
\ee
In the Miwa variables, it is $\tau^{KP}(X)=\sum_\lambda \pi_\lambda s_\lambda(\pb(X))=\sum_\lambda \pi_\lambda s_\lambda(X)$.
Here the set of coefficients $\pi_\lambda$ defines a solution to the KP
equation. These coefficients solve special bilinear relations: the Pl\"ucker relations
for the Pl\"ucker coordinates of a chosen point of the Sato Grassmannian \cite{Sato1,Sato2,Sato3},\cite{J1,J2,JM}.
Each point of the Sato Grassmannian yields a solution to the infinite set of the higher
KP equations (the KP hierarchy) and to the KP equation itself\footnote{
If one thinks about the
KP equation as the (2+1)-dimensional equation of waves in ocean, then,
roughly speaking, a point of the infinite Sato Grassmannian can be viewed as an infinite matrix
which defines the initial configuration of the surface profile which is a function of two variables.}.

The UC hierarchy and the UC tau functions were constructed in \cite{Tsuda}.
It depends on two sets of times, say $\pb=(p_1,p_2,\dots)$ and $\tilde\pb=(\tilde{p}_1,\tilde{p}_2,\dots)$.
Each UC tau function can
be defined in terms of a selected pair of KP tau function as follows.
Consider an independent pair of KP tau functions in the Sato form
\be
\tilde\tau^{KP}(\tilde\pb)=\sum_\lambda \tilde\pi_\lambda s_\lambda(\tilde\pb),\quad
\tau^{KP}(\pb)=\sum_\lambda \pi_\lambda s_\lambda(\pb),
\ee
where $\tilde\pi_\lambda$ and $\pi_\lambda$ are the Pl\"ucker coordinates of a pair of points in the Sato Grassmannian
\cite{Sato1}, \cite{JM}
 which specify these $\tilde\tau^{KP}$ and $\tau^{KP}$.

According to \cite{Tsuda}, any UC tau function can be defined as
\be\label{piUCpi}
\tau^{UC}(\tilde\pb,\pb)=\tilde\tau^{KP}(\tilde\pb-\partial)\cdot \tau^{KP} (\pb)=
\tilde\tau^{KP}(\pb-\tilde\partial)\cdot \tilde\tau^{KP}(\tilde\pb)=
\ee
\be\label{piUCpi'}
=\sum_{\lambda,\mu} \tilde\pi_\lambda \pi_\mu s_{[\lambda,\mu]}(\tilde\pb,\pb),
\ee
or, the same $\tau^{UC}(\tilde\pb,\pb)=
[ \tilde\tau^{KP}(\tilde\pb)\ast \tau^{KP} (\pb) ]_1$.

If, for $\ell(\lambda)>N$, the Pl\"ucker coordinates $\pi_\lambda,\tilde\pi_\lambda=0$, then  we can write
$\tau^{UC}(\tilde\pb,\pb)=
[ \tilde\tau^{KP}(\tilde\pb)\ast \tau^{KP} (\pb) ]_2$.

If the KP tau functions
\be
\tilde\tau^{KP}(X_1)=\sum_\lambda \tilde\pi_\lambda s_\lambda(X_1),\quad
\tau^{KP}(X_2)=\sum_\lambda \pi_\lambda s_\lambda(X_2),
\ee
are written in the Miwa variables given by eigenvalues of matrices $X_1$ and
$X_2$ of the form (\ref{block}), and, for $\ell(\lambda)>n$, the Pl\"ucker coordinates $\pi_\lambda,\tilde\pi_\lambda=0$, then
\be
\tau^{UC}(X_1,X_2)=
[ \tilde\tau^{KP}(X_1)\ast \tau^{KP} (X_2) ]_3.
\ee

The Schur functions are the simplest polynomial KP tau function;
choosing $s_\lambda(\tilde\pb)$ and $s_\mu(\pb)$ as the KP tau functions, one gets
the simplest
polynomial UC tau function $s_{[\lambda,\mu]}(\tilde\pb,\pb)$.

Any KP tau function solves the KP Hirota bilinear equation \cite{JM}, which is a difference
equation involving an arbitrary shift of KP higher times \cite{JM}, and, because of the grading of the power sum
variables, this equation results in the infinite number
of differential equations with respect to the variables $p_1,p_2,\dots$
commonly called KP hierarchy.

Similarly, any UC tau function solves a pair of UC Hirota bilinear difference equations
with respect to a pair of sets of power sum variables \cite{Tsuda} denoted as
$\tilde\pb,\pb$ in (\ref{piUCpi}). However,
as soon as $p_m$ and $\tilde p_m$ have opposite gradings, in general, these difference
Hirota bilinear equations result in differential equations with infinitely many terms (see Appendix B).

The UC tau functions have applications in study of the Painlev\'e equations \cite{TsudaPainleve},
of q-difference equations \cite{TsudaDiff}, of
isomonodromic deformations \cite{TsudaIsomon},
in knot theory \cite{LMO1} and in topological strings \cite{LMO1}. Both in \cite{LMO1} and in the present paper, we
discuss UC tau functions that are partition functions of various models of random matrices (matrix models).

\paragraph{Relations of the UC hierarchy to matrix models.}
We are going to realize the UC tau functions (\ref{piUCpi})
as certain integrals over matrices. The list of matrix models as they were known in 1960-ies
is presented in the famous textbook \cite{Mehta}, and now there is an ample set of interesting
matrix models under study. Here we enlarge this set of solvable matrix models
simultaneously producing more examples of explicit UC tau functions.

There are few ways to relate the two topics. They are based on different choices of the mating
$[ f\ast  g]_i,\, i=0,1,2$, see below.

\paragraph{Method I.}
 The most straightforward relation of the UC hierarchy to matrix models
is to use the definition (\ref{piUCpi}), where
at least one tau function is a matrix integral. There is a rich list of matrix integrals, see \cite{GMMMO}-\cite{AntOV}. We describe this way in more detail elsewhere. Here we just present two examples:

{\bf Example 1.} Take a pair of unitary matrix integrals \cite{PS1,PS2,BMS,KhM} with couplings
$\pb_+^{(i)}=\left(p_1^{(i)},p_2^{(i)},\dots  \right)$ and
$\pb_-^{(i)}=\left(p_{-1}^{(i)},p_{-2}^{(i)},\dots  \right)$:
\be
\tau_i(\pb_+^{(i)},\pb_-^{(i)})=\int_{\mathbb{U}_{N_i}}
e^{\sum_{m\neq 0} \frac 1m p^{(i)}_m\tr U_i^m} d_*U_i=\sum_{\lambda\atop \ell(\lambda)\le N_i}
s_\lambda(\pb_+^{(i)})s_\lambda(\pb_-^{(i)}),\quad i=1,2,
\ee
where $d_*U_i$ denotes the Haar measure on the unitary group $\mathbb{U}_{N_i}$ (see Appendix A), and
the sum over partitions $\lambda$ is restricted by their length (the length $\ell(\lambda)$
of $\lambda=(\lambda_1,\dots,\lambda_n),\,\lambda_n\neq 0$ is $n$),
 and produces a UC tau function
according to (\ref{piUCpi}) by pairing, say, $\pb^{(1)}_-$ and $\pb^{(2)}_-$. We obtain
\be\label{ex1}
\tau^{UC}\left(\pb_-^{(1)},\pb_-^{(2)}|\pb_+^{(1)},\pb_+^{(2)}\right)=
\int_{\mathbb{U}_{N_1}} \int_{\mathbb{U}_{N_2}}\det\left(1-U_1\otimes U_2\right)\prod_{i=1,2}
e^{\sum_{m\neq 0} \frac 1m p^{(i)}_m\tr U_i^m} d_*U_i =
\ee
$$
=\sum_{\ell(\lambda),\ell(\mu)\le\min(N_1,N_2)}
s_\lambda(\pb_+^{(1)})s_{[\lambda,\mu]}(\pb_-^{(1)},\pb_-^{(2)})s_\mu(\pb_+^{(2)}).
$$
In (\ref{ex1}), the sets $\pb^{(1)}_-$ and $\pb^{(2)}_-$ play the role of UC times,
while the sets $\pb^{(1)}_+$ and $\pb^{(2)}_+$ are parameters.

{\bf Example 2.} For $\tau^{KP}_1(\tilde\pb)$, one takes the simplest one, $\tilde\pb=(\tilde p,0,0,\dots)$, and, for
$\tau^{KP}_2(\pb)$, one takes, say, the well-known one-matrix model \cite{GMMMO} where
$\pb=(p_1,p_2,\dots)$ are coupling constants. One obtains
\be
\tau^{UC}(\pb,\tilde\pb)=\int \left(\tilde p -\tr M \right)e^{\sum_{m>0} \frac1m p_m\tr M^m} dM,
\ee
where the measure $dM$ is the standard measure on Hermitian matrices \cite{Mehta}.
 One can treat this method as an application of the mating.

\paragraph{Method II.}
It is described in \cite{LMO1},
and is based on the relation
\be\label{mating1}
\left[s_\lambda(\pb) \ast s_\mu(\pb)  \right]_2:=
\ee
\be\label{Mir2}
\int_{\mathbb{U}_N}
s_\lambda\left(\pb^{(1)}+\pb(U)\right)s_\mu\left(\pb^{(2)}-\pb(U^\dag)\right)d_*U=
\sum_{\nu}(-1)^{|\nu|}s_{\lambda/\nu}(\pb^{(1)})s_{\mu/\nu^t}(\pb^{(2)})
\ee
\be\label{Mir3}
=s_{[\lambda,\mu]}(\pb^{(1)},\pb^{(2)}),
\ee
where $\ell(\lambda),\ell(\mu)\le N $.

One can use different matrices instead of the unitary ones if there exists a counterpart of
(\ref{mating1})-(\ref{Mir2}) equal to the universal character (\ref{Mir3}).

\paragraph{Method III.}
This way needs a certain generalization of the UC hierarchy (``enlarged UC hierarchy'') where the modified UC tau function can not be presented in form (\ref{piUCpi}). This is the subject of \cite{LMO3}.

\paragraph{Method IV.}
 This is the way we describe in the present paper. It is more tricky, and involves
a pair of matrix models defined on a graph which depend on source matrices placed to the vertices
of the graph, see \cite{NO-TMP},\cite{2DYM},\cite{AOV}.

This way is based on the UC mating of the Schur functions of the matrix argument
\be\label{mating2'}
\left[ s_\lambda(A_1) \ast s_\mu(\tilde A_2)\right]_3,
\ee
see below.

This paper mainly considers the relationships between matrix models whose partition functions are expressed through universal characters given by (\ref{mating2'}) and UC tau functions.

\section{Matrix models associated with the product of two spheres with embedded graphs}

\subsection{General description}

Partition function of many matrix models can be written as a sum over partitions like
$$
Z(\pb,\pb(X)|\dots) = \sum_\lambda (\dots) s_\lambda(\pb)s_\lambda(\pb(X)),
$$
where $\pb(X)=\left(\tr X, \tr X^2,\dots\right)$, $X\in {\rm Mat}_{N\times N}$.
The family of such matrix models is presented below in (\ref{I}) and (\ref{II})
where the matrix model is associated with an embedded graph on an orientable surface.
For $Z$ to be a two-component KP tau function (where $\pb$ and $\pb(A)$ are the time variables),
the graph must be drawn on a sphere.
Dots in the argument of $Z$ denote additional parameters.
Here, the model depends on the set $\pb$ which we call coupling
constants and a matrix $A$ which we call a source matrix.
For a chosen pair of matrix models (each is defined by a graph drawn on $\mathbb{S}^2$) one can mate them to get new matrix model. As we already mentioned, there are different ways of mating.
For instance,
$$
Z^{UC}(\pb^1,\pb^2,\pb(X_1),\pb(X_2)|\dots)=
$$
$$
=[ Z_1(\pb^1,\pb(X_1)|\dots)\ast
Z_1(\pb^2,\pb(X_2)|\dots)]_i =\sum_{\lambda,\mu} (\dots)_i s_\lambda(\pb(X_1))
s_{[\lambda,\mu]}(\pb^1,\pb^2) s_\mu(\pb(X_2)),\quad i=1,2.
$$
This is a plan how to construct new matrix model, details depends on the choice of
$Z_1$ and $Z_2$ and the choice of mating. The realization is rather straightforward,
see (\ref{ex1})
as the example and in \cite{LMO1} we wrote down an example of mating (\ref{mating1}).

One can consider
$$
Z^{UC}(\pb^1,\pb^2,\pb(X_1),\pb(X_2)|\dots)=[ Z_1(\pb^1,\pb(X_1)|\dots)\ast
Z_1(\pb^2,\pb(X_2)|\dots)]_3=
$$
$$
=\sum_{\lambda,\mu} (\dots) s_\lambda(\pb^1)
s^{[n]}_{[\lambda,\mu]}(\pb(X_1),\pb(X_2)) s_\mu(\pb^2),
$$
based on mating (\ref{mating2'}) with certain $n$.

We consider two examples below.

In works \cite{NO-TMP},\cite{2DYM},\cite{AOV}, matrix models on embedded graphs on Riemann surfaces (graphs with all faces homeomorphic to disks) were proposed. The corresponding matrix integrals are calculated in terms of the Schur functions of coupling constants attached to the graph faces and of matrices attached to the graph vertices. These matrix integrals defined by graphs on a sphere under certain restrictions on coupling constants and on matrices at vertices can be made equal to the tau function of the KP hierarchy or to the tau function of a two-component KP hierarchy.

We start with examples.

\subsection{Examples. Integrals over unitary matrices}

\subsubsection{Example 1 \label{Ex1}}

Consider the following 4-matrix model
\be\label{J}
J(X_1,X_2,\pb^{(1)},\pb^{(2)},\tilde\pb^{(1)},\tilde\pb^{(2)})=\int_{\mathbb{U}_n\times \mathbb{U}_n}
d_*U_1d_*U_2 \,{\cal G}(U_1,U_2)
\prod_{a=1,2}
\int_{\mathbb{U}_N}d_*V_a \exp W(V_a,U_a,|X_a,\pb^{(a)},\tilde\pb^{(a)}),
\ee
where $N>n$, where
\be
{\cal G}(U_1,U_2)=\det(I_n\otimes I_n-U_1\otimes U_2),
\ee
\be
W(V_a,U_a|X_a,\pb^{(a)},\tilde\pb^{(a)})=
\sum_{m>0}\left(\frac 1m p^{(a)}_m
\tr \left((V_aC_a)^m\right) + \frac 1m \tilde{p}^{(a)}_m
\tr \left((V_a^\dag C_{-a})^m\right) \right),
\ee
and where the matrices $C_{\pm a}$ are restricted to have the following block
structure for both $C_1C_{-1}$ and $C_2C_{-2}$:
\be\label{CaC_a}
\left(C_aC_{-a}\right)_{i,j}=\begin{cases}
                              (U^\dag_a)_{i,j}\quad i,j \le n \\
                            (X_a)_{i,j}\quad n<i,j\le N-n \\
                            0\quad {\rm otherwise}.
                             \end{cases}
\ee
We denote by $\pb$ the four sets $\pb^{(a)}=(p^{(a)}_1,p^{(a)}_2,\dots)$, $\tilde\pb^{(a)}=(\tilde{p}^{(a)}_1,\tilde{p}^{(a)}_2,\dots)$, $a=1,2$, and the two matrices $X_a$, $a=1,2$ play the role of coupling constants of the model.
Here $d_*U$ (and $d_*V$) is the Haar measure on the unitary group (see (\ref{Haar}) in Appendix A).

Integral (\ref{J}) can be evaluated in an explicit way.
First,
we recall that the Schur function $s_\lambda(X),\,X\in {\mathbb{GL}}_N$ is the character of the polynomial representation of the linear group labeled by the partition $\lambda$. A pair of
partitions labels the rational representation of the linear group, and the related character
is called universal character \cite{Koike} and is denoted by $s_{[\lambda,\mu]}(X_1,X_2)$. The both
characters can be rewritten in terms of the power sum variables. That is, if
\be
p_n=p_n(X)=\tr \left(X^n\right),
\ee
we can write $s_\lambda(X)=s_\lambda(\pb(X))$. Similarly, for
\be\label{pX1pX2}
p_n=p_n(X_1)=\tr \left(X_1^n\right),\quad
p^*_n=p^*_n(X_2)=\tr \left(X_2^n\right),
\ee
we write $s_{[\lambda,\mu]}(X_1,X_2)=s_{[\lambda,\mu]}(\pb,\pb^*)$.

\br
Formula (\ref{pX1pX2}) implies that $s_{[\lambda,\mu]}(X_1,X_2)=0$ in case $\ell(\lambda)>
{\rm rank} X_1$ and in case $\ell(\mu)>{\rm rank} X_2$.
\er

We have
\bp\label{Prop1}

\be\label{character-expansion}
J(X_1,X_2,\pb)=\sum_{\lambda,\mu\atop\ell(\lambda),\ell(\mu)\le N}
\frac{s_\lambda(\pb^{(1)})s_\lambda(\tilde\pb^{(1)}) }{s_\lambda(I_N)}
s^{[n]}_{[\lambda,\mu]}(X_1,X_2)
\frac{s_\mu(\pb^{(2)})s_\mu(\tilde\pb^{(2)}) }{s_\mu(I_N)},
\ee
where $s_\lambda(I_N):=s_\lambda(\pb(I_N))$ is known to be dimension of the irreducible
representation $\rho_\lambda$ of the linear group $\mathbb{GL}_N$ lebelled by $\lambda$ \cite{Mac}, which can be written
down in an explicit way in terms of the partition $\lambda$ :
\be
\dim_{\mathbb{GL}_N}\rho_\lambda=s_\lambda(I_N)=\frac{\prod_{i<j\le N}(\lambda_i-\lambda_j-i+j) }{\prod_{i=1}^N (\lambda_i-i+N)!} \prod_{(i,j)\in\lambda}(N+j-i).
\ee
\ep

The proof of Proposition \ref{Prop1} follows from one of the definitions of the universal character
given by formula (\ref{def2})
 and from the following set of lemmas:

\bl\label{lemma0}
\be
e^{\sum_{m>0}\frac 1m p_m \tr(A^m)}=\sum_\lambda  s_\lambda(\pb)s_\lambda(A),
\ee
\el
which is the Cauchy-Littlewood identity \cite{Mac}. We apply it to the integrand of (\ref{J}).

\bl\label{lemma1'}
\be\label{integral1}
\int_{\mathbb{U}_k} s_\lambda(UA)s_\mu(U^\dag B)d_*U=\delta_{\lambda,\mu}\frac{s_\lambda(AB)}{s_\lambda(I_k)} ,
\ee
\el
which is known;  see \cite{Mac}.

This integral is associated with the graph which is a loop drawn on the sphere: one edge, one
vertex and two faces (this can be seen as equator on the globe with a vertex on the equator), see Section \ref{graphs} and Fig 1b in Appendix \ref{graphs-for-MM}.

\bl\label{lemma2}
\be
\det\left(I_n-U_1 \otimes U_2  \right)=\sum_{\nu} (-1)^{|\nu|}s_\nu(U_1)s_{\nu^t}(U_2).
\ee
\el
which is also a version of the Cauchy-Littlewood identity
$\prod_{i,j}(1-x_iy_j)=\sum_{\nu} (-1)^{|\nu|}s_\nu(x) s_{\nu^t}(y)$, see \cite{Mac} Page 35.

\bl\label{lemma3}
\be
s_\lambda(C_a C_{-a})=\sum_\nu s_{\lambda/\nu}(X_i)s_\nu(U^\dag_i).
\ee
\el
which follows from the fact that $t_n:=\tr(C_a C_{-a})^n = t^{(1)}_n+t^{(2)}_n$,
where $t^{(1)}_n=\tr(X_a)^n$, $t^{(2)}_n=\tr(U^\dag_a)^n$ and from
$s_\lambda(t^{(1)}_n+t^{(2)}_n)=\sum_\nu s_{\lambda/\nu}(t^{(1)}_n) s_\nu (t^{(2)}_n)$.

Further, we want to relate the right hand side of (\ref{character-expansion}) to a tau function.
To do it, we need to have both $
\frac{s_\mu(\pb^{(i)})s_\mu(\tilde\pb^{(i)}) }{s_\mu(I_N)},\, i=1,2$ to be Pl\"ucker
coordinates $\pi^{(i)}_\lambda$.

Let $r$ be any function on the lattice $\mathbb{Z}$, and
$\pb$ is any set of power sum variables.
By direct computation, one can prove
that the following function of $\lambda$ written in the form $s_\lambda(\pb)\cdot\prod_{(i,j)\in\lambda} r(j-i)$  solves the Pl\"ucker equations.

Notice that if one chooses
\be\label{p(c)}
\tilde{\pb}^{(i)}=\tilde{\pb}^{(i)}(c_i,z_i),\quad \tilde{p}^{(i)}_m(c_i,z_i)=c_i z_i^m,
\quad i=1,2
\ee
then
\be
e^{\tilde{W}_i(\pb(c_i,z_i))}=\det\left(I_N-z_iV^\dag C_{-1}\right)^{-c_i}.
\ee
We also know (see \cite{Mac}) that, for any $c$ and $z$, there is the following equality:
\be
s_\lambda(\pb(c,z))=(c)_\lambda z^{|\lambda|}s_\lambda(\pb_\infty),\quad
\pb_\infty:=(1,0,0,\dots),
\ee
where
\be\label{content}
(c)_\lambda=\prod_{(i,j)\in\lambda}(c+j-i),
\ee
is called the content product. It is important for us that
\be\label{Plucker-example}
\frac{c_\lambda}{(N)_\lambda} s_\lambda(\tb)=\pi_\lambda
\ee
for any $c$ and for any set $\tb$
solves the Pl\"ucker equation (see Appendix B).
Suppose $c\le n$ is an integer. In this case
in (\ref{Plucker-example}), $\pi_\lambda=0$ for $\ell(\lambda)>n$.

Note that we also have
\be
s_\lambda(I_N)=(N)_\lambda s_\lambda(\pb_\infty).
\ee
(see \cite{Mac}).

Then, by Lemma \ref{mate-2-UC}, we come to the following

\bp\label{Prop2}
Let both $c_i\le n$ be integers, and $z_i\in\mathbb{C}$ (where $i=1,2$).
Then the matrix integral (\ref{J}) is the tau function $\tau^{UC}(\pb,\pb^*)$ of the UC hierarchy, introduced in \cite{Tsuda}
with the UC time variables
\be
p_m=\tr \left((X_1)^m\right),\quad p^*_m=\tr \left((X_2)^m\right),
\ee
namely, one can write
\be
J(\pb(X_1),\pb^*(X_2),\pb^{(1)},\pb^{(2)})=\sum_{\lambda,\mu\atop\ell(\lambda),\ell(\mu)\le N}
  \pi^{(1)}_\lambda(\pb^{(1)})
s_{[\lambda,\mu]}(X_1),X_2)
\pi^{(2)}_\mu(\pb^{(2)}) =
\ee
$$
= \tau^{UC}(\pb(X_1),\pb^*(X_2)|c_1,c_2,z_1,z_2,\pb^{(1)},\pb^{(2)})
$$
where $c_1,c_2,z_1,z_2,\pb^{(1)},\pb^{(2)}$ are free parameters, and
where both
\be
\pi^{(1)}_\lambda(\pb^{(1)}) =z_1^{|z|}\frac{(c_1)_\lambda}{(N)_\lambda} s_\lambda(\pb^{(1)}),  \quad {\rm and} \quad
\pi^{(2)}_\mu(\pb^{(2)})=s_\mu(\pb^{(2)}) z_2^{|\mu|}\frac{(c_2)_\mu}{(N)_\mu},
\ee
solve the Pl\"ucker relation for the KP Grassmannian.
\ep

\subsubsection{Example 2 \label{Ex2}}

Consider the following 4-matrix model
\be\label{J2}
I(X_1,X_2,C_1,C_2,\pb^{(1)},\pb^{(2)})=\int_{\mathbb{U}_n\times \mathbb{U}_n}
d_*U_1d_*U_2 \,{\cal G}(U_1,U_2)
\prod_{a=1,2}
\int_{\mathbb{U}_N}d_*V_a \exp W(V_a,U_a,|X_a,\pb^{(a)}),
\ee
where $N>n$, where
\be
W(V_a,U_a|X_a,\pb^{(a)})=
\sum_{m>0} \frac 1m p^{(a)}_m
\tr \left((V_aC_aV_a^\dag C_{-a})^m\right),
\ee
and where the both matrices $C_{- a},\,a>0$ are restricted to have the following block
structure:
\be\label{Ca}
\left(C_{-a}\right)_{i,j}=\begin{cases}
                              (U^\dag_a)_{i,j}\quad i,j \le n, \\
                            (X_a)_{i,j}\quad n<i,j\le N-n, \\
                            0\quad {\rm otherwise},
                             \end{cases}
\ee
while $C_{a},\,a>0$ are still arbitrary.
The two matrices $C_{a},\,a>0$, the two sets $\pb^{(a)}=(p^{(a)}_1,p^{(a)}_2,\dots)$ which we denote by $\pb$ and the two matrices $X_a$ play the role of coupling constants of the model.

Integral (\ref{J2}) can be explicitly evaluated:
\bp\label{Prop3}

\be\label{character-expansionJ2}
I(X_1,X_2,C_1,C_2,\pb^{(1)},\pb^{(2)})=\sum_{\lambda,\mu\atop\ell(\lambda),\ell(\mu)\le N}
\frac{s_\lambda(\pb^{(1)})s_\lambda(C_1) }{s_\lambda(I_N)}
s^{[n]}_{[\lambda,\mu]}(X_1,X_2)
\frac{s_\mu(\pb^{(2)})s_\mu(C_2) }{s_\mu(I_N)}.
\ee
\ep

The proof is identical to the previous case, however, instead of Lemma \ref{lemma1'},
we use

\bl\label{lemma1a}
\be
\int_{\mathbb{U}_k} s_\lambda(UAU^\dag B) d_*U=\frac{s_\lambda(A)s_\lambda(B)}{s_\lambda(I_k)}.
\ee
see \cite{Mac}.
\el

This integral is associated with the graph that is a segment drawn on the sphere: one edge, two
vertices and one face, see Section \ref{graphs} and Fig 1c in Appendix
\ref{graphs-for-MM}.

Further, we want to relate the right hand side of (\ref{character-expansionJ2}) to a tau function.
It is a UC tau function in the case of the both $\frac{s_\lambda(\pb^{(i)})s_\lambda(C_i) }{s_\lambda(I_N)}$
equal to Pl\"ucker coordinates, say, to $\pi^{(i)}_\lambda,\,i=1,2$, and, in addition,
we get $\pi^{(i)}_\lambda=0$ for $\ell(\lambda)>n$.

The functions $\frac{s_\lambda(\pb^{(i)})s_\lambda(C_i) }{s_\lambda(I_N)}$ on the set of partitions $\lambda$ can be equal to Pl\"ucker coordinates in the following possible cases:

\begin{itemize}
\item[(i)] either $\pb^{(i)}$ is of the form (\ref{p(c)}) (which provides that the set $\pi^{(i)}_\lambda$
for all $\lambda$ is a product of the Schur function $s_\lambda(C_i)$ times a content product,
namely
\be\label{op1}
\pi^{(i)}_\lambda = s_\lambda(C_i)z_i^{|\lambda|}\prod_{(k,j)\in \lambda}\frac{c_i+j-k}{N+j-k}
\ee
which, as one can check, solves the Pl\"ucker relations. To provide the condition
$\pi^{(i)}_\lambda=0$ for $\ell(\lambda)>n$, one needs either (a) $c_i\le n$ to be integer, or (b) ${\rm rank}\ C_i\le n$, with $c_i$  being an arbitrary complex number.

\item[(ii)] or, the spectrum of $C_i$ consists of $n_i$ units and $N-n_i$ zeroes. In this case,
\be\label{op2}
\pi^{(i)}_\lambda = s_\lambda(\pb^{(i)})\prod_{(k,j)\in\lambda} \frac{n_i+j-k}{N+j-k}
\ee
which has the same form as in the previous case (\ref{op1}) and similarly solves the Pl\"ucker relation. To provide the condition
$\pi^{(i)}_\lambda=0$ for $\ell(\lambda)>n$, one needs either (a) $n_i\le n$, or
(b) the rank of $C_i$ is arbitrary and there exists a matrix $Y$ with rank does not exceeding
$n$ such that $\pb^{(i)}=\pb(Y)$. In the latter case, $s_\lambda(\pb^{(i)}(Y))=0$ for $\ell(\lambda)>n$. Thus, in the both cases (a) and (b), $\pi^{(i)}_\lambda=0$ for $\ell(\lambda)>n$.
\end{itemize}

There are four possibilities of achieving
\be
\pi^{(i)}_\lambda =0\quad {\rm if}\quad \ell(\lambda)> n,
\ee
namely:

\begin{itemize}
\item[(a)]

\be\label{op1'}
\pi^{(i)}_\lambda = s_\lambda(C_i)z_i^{|\lambda|}\prod_{(k,j)\in \lambda}\frac{c_i+j-k}{N+j-k}
\ee
where $c_i$ is integer and $c_i\le n$.

\item[(b)]

\be\label{op1''}
\pi^{(i)}_\lambda = s_\lambda(C_i)z_i^{|\lambda|}\prod_{(k,j)\in \lambda}\frac{c_i+j-k}{N+j-k}
\ee
where $c_i$ is arbitrary, and ${\rm rank}\, C_i\le n$.

\item[(c)]

\be\label{op2'}
\pi^{(i)}_\lambda = s_\lambda(\pb^{(i)})\prod_{(k,j)\in\lambda} \frac{n_i+j-k}{N+j-k}
\ee
where $n_i\le n$ ($n_i$ is the rank of $C_i$).

\item[(d)] There exists a matrix $Y$ such that $\pb^{(i)}=\pb(Y)$ (see the notation (\ref{powerSums}))

\be\label{op2''}
\pi^{(i)}_\lambda = s_\lambda(\pb^{(i)}(Y))\prod_{(k,j)\in\lambda} \frac{n_i+j-k}{N+j-k}
\ee
where $n_i$ is arbitrary, and the rank of $Y$ does not exceed $n$.
\end{itemize}

These are equalities imposed to the parameters $C_1,C_2,\pb^{(1)},\pb^{(2)}$ to make the integral
$I$ a UC tau function in the Miwa variables $X_1,X_2$.

\bp\label{Prop4}

If each of $\pi^{(i)}_\lambda= \frac{s_\lambda(\pb^{(i)})s_\lambda(C_i) }{s_\lambda(I_N)},\,i=1,2$
satisfies any of the conditions (\ref{op1'})-(\ref{op2''}), then the integral
(\ref{character-expansionJ2}) can be identified as the UC tau function
\be
I(X_1,X_2,C_1,C_2,\pb^{(1)},\pb^{(2)})=\sum_{\lambda,\mu}
\pi^{(1)}_\lambda(C_1,\pb^{(1)}) s_{[\lambda,\mu]}(X_1,X_2)\pi^{(2)}_\mu(C_2,\pb^{(2)})=
\ee
\be
=\tau^{UC}(X_1,X_2|C_1,C_2,\pb^{(1)},\pb^{(2)})
\ee
in the Miwa variables $X_1,X_2$ (or, the same, in the eigenvalues of the matrices $X_1$ and $X_2$).

\ep

Proposition \ref{Prop4} provides sufficient (but not necessary) conditions for the integral
(\ref{character-expansionJ2}) to be identified as a UC tau function.

Both examples considered above are related to the mating of two matrix models each of which was actually
associated with a graph. In the Example 1 both matrix models can be related to the Fig 1b. In the Example 2
both models which we mate can be realted to the Fig 1c. In the next section, we recall the construction
of matrix models related to an arbitrary (embedded) graph presented in
\cite{NO2020},\cite{NO-TMP},\cite{AOV} and the mating of such models under certain conditions results in the  UC tau functions.

\section{Multi-matrix models of complex and unitary matrices associated with embedded graphs \label{graphs}}

Here we present a general construction of matrix models on graphs with corner matrices (or the same on graphs with source matrices). Corner matrices are free parameters and allows to marry two different models on two different graphs.

From previous studies \cite{AOV} we know that to obtain a multimatrix model whose partition function can be identified with the tau function KP (and TL), our graph must be plotted on a sphere.
Thus, we link two different matrix models corresponding to different graphs on two spheres. And we mate these models, their partition functions and tau functions, obtaining the tau function UC at the output.

\paragraph{Embedded graphs.}

An embedded graph put on an oriented Riemann surface without boundary has only faces that are homeomorphic to a disc. The boundary of each face
consists of the sides of ribbon edges. One assigns the positive (counterclockwise) orientation
to the boundary of each face. It means that each ribbon edge consists of oppositely
directed arrows. 
Thus, the boundary of the face consists of arrows successively placed next to each other with a positively chosen orientation.
Let us number each edge with positive numbers from 1 to $n$, where $n$ is the total number of edges. In the case of sphere, the graph with $F$ faces possesses $V=2+n-F$ vertices. To the sides of edge with number $i$ ($i>0$), we assign numbers $i$ and $-i$, thus, all sides of the graph are numbered
with numbers from the set $\pm 1,\dots, \pm n$. Let us denote this set as $\mathfrak{N}$, with
${\rm card}\, \mathfrak{N}=2n$. Let us also number the faces
of the graph with $1,\dots,F$.
By going around each face $f_i$ in the positive direction, one obtains a set of numbers from the list above, and such a set is defined up to a cyclic permutation; denote such a set
by $\mathfrak{f}_i$ and call it {\it face cycle} associated with $f_i$.

\paragraph{Combinatorial description of embedded graphs} can be found, say, in \cite{ZL}, section 1.3.3. We send reader to this wonderful textbook for details.
We suggest you take a look at Appendix D for illustrative examples.
Let us consider the symmetric group $S_{2n}$, which acts on $\mathfrak{N}$.
 Consider the element of $S_{2n}$, which is the product of the face cycles:
 \be\label{face-cycle}
 (\mathfrak{f}_1)\cdots (\mathfrak{f}_F) \in S_{2n}\,,
 \ee
 where the order of the factors is inessential because, by construction, these are non-intersecting
 cycles.

 Let us number the vertices of the graph with $1,\dots,V$.
 To each vertex $i$, we assign a vertex cycle which is an ordered set of numbers
associated with the out-coming arrows obtained when passing clockwise around the vertex.
 (We recall that these are the numbers from the set $\mathfrak{N}$.) This way one obtains a cycle
 of $S_{2n}$. We denote it as $\mathfrak{v}_i$. As a result, one obtains
 the following product of non-intersecting cycles:
$$
(\mathfrak{v}_1)\cdots (\mathfrak{v}_V) \in S_{2n}\,.
$$

Between the two described elements of $S_{2n}$, there is a well-known relation associated with any embedded graph:
\be\label{graph-combinatorial}
\sigma \circ (\mathfrak{f}_1)\cdots (\mathfrak{f}_F) =(\mathfrak{v}_1)\cdots (\mathfrak{v}_V),
\ee
where, at the left-hand side of this formula, there is a composition of the involution $\sigma^2=1$ without
fixed points and the product of face cycles. In our notations, $\sigma$ acts as simultaneous
transpositions $i \leftrightarrow -i$ for all $i=1,\dots,n$:
$$
\sigma=(1,-1)(2,-2)\cdots (n,-n)
$$
which the element of $S_{2n}$ with cycle structure $(2^n)$.

The simplest example $n=2$ is considered in Appendix \ref{graphs-for-MM} and is related to Fig 1b and Fig 1c.

\br

To compute the right-hand side of \ref{graph-combinatorial} using the transposition action on the product of cycles on the left-hand side, recall the cut-and-join action
of transposition on the product of cycles. A couple of simple examples: (i) if we take $n=1$ (one edge) and one cycle ${\mathfrak f}=(1,-1)$, we get the ``cut'' action of transposition:
$(1,-1)(1,-1)=(1)(-1)={ \mathfrak v}_1{\mathfrak v}_2$. The only graph with one edge, one face, and two
vertices is the segment, denoted by $\Gamma_1$. (ii) Next, let's take $n=1$, two cycles
${\mathfrak f}_1=(1)$ and ${\mathfrak f}=(-1)$, and get the "joining" action: $(1,-1)(1)(-1)=(1,-1)={\mathfrak v}$.
This is the graph $\Gamma_2$, which can be represented as the equator on a sphere with a vertex
on the equator, $\Gamma_2$ is the dual graph of $\Gamma_1$.

For completeness, let's give another example, a little more complicated:
we want to describe the graph defined by cycles $a$ and $b$ on the torus. To do this, we consider $n=2$ (two ribbon edges) and the cycle ${\mathfrak f }=(1,2,-1,-2)$. Then we first apply transpostion $(2,-2)$,
breaking ${\mathfrak f}$ into a pair of cycles, and then we apply $(1,-1)$ , which will merge these two new cycles into one. We get the cycle $(1,-2,-1,2)={\mathfrak v}$, which is a vertex cycle.

\er

Since $\sigma^2=1$, we additionaly get
\be\label{graph-combinatorial*}
\sigma \circ (\mathfrak{v}_1)\cdots (\mathfrak{v}_V) =(\mathfrak{f}_1)\cdots (\mathfrak{f}_F).
\ee
Formulas (\ref{graph-combinatorial}) and (\ref{graph-combinatorial*}) are related to dual graphs.

This is a combinatorial description of the embedded graph\footnote{
In \cite{ZL}, it is written in a slightly different way in terms of numbering the half-edges of the dual graph.} drawn on a surface.

As it was noticed in \cite{NO-TMP}, we have a wonderful counterpart of these relations in terms
of integrals over matrices.

To construct matrix models together with their statistical sum, one can either use the relation
(\ref{graph-combinatorial}), forgetting about the graphs (combinatorial approach), or forget about the permutation group and use visualization with graphs (geometric approach). It is a matter of preference. In the combinatorial approach we start from the element in $S_{2n}$ with the cycle structure $(\mathfrak{f}_1)\cdots (\mathfrak{f}_F)$ which defines the matrix model, and then thanks to (\ref{graph-combinatorial}) we find $(\mathfrak{v}_1)\cdots (\mathfrak{v}_V)$ which defines
the partition function.

The way to relate (\ref{graph-combinatorial}) to matrix models is to send $i$ from the set $\{\pm 1,\dots,\pm n\}$ to $C_i\in {\rm Mat}_{N\times N}$, and ${\mathfrak f}_a\to {\cal F}_a,\,a=1,\dots,F$, $
{\mathfrak v}_b\to {\cal V}_b,\, b=1,\dots,V$, see (\ref{face-cycle-product}), (\ref{dressed-face-cycle product}) and (\ref{vertex-cycle-product}) below.

Let us remark that the first example of such multi-matrix models with the ensemble of unitary matrices was suggested in \cite{Witten} within the context of 2D Yang-Mills
theory. However, there were no source matrices, which are crucial for our paper, and
there were no links with tau functions and no links with the property (\ref{graph-combinatorial}) below. See also similar works  \cite{Alexandrov}.

\paragraph{Multi-matrix Ginibre ensembles, ensembles of unitary matrices, mixed ensembles.\label{mixed}}

The complex Ginibre ensemble \cite{Gin} is defined by the measure on the space of complex $N\times N$ matrices:
\be\label{GUEmeasure}
d\mu(Z)= C e^{-N\Tr ZZ^\dag}\prod_{i,j\le N} d^2 Z_{i,j}\,.\quad \int d\mu(Z)=1
\ee
The $n_1$-matrix complex Ginibre ensemble is defined by the measure $d\mu(Z_1,\dots,Z_{n_1})=\prod_{i=1}^{n_1} d\mu(Z_i)$ on the space of $N\times N$ matrices
$Z_1,\dots,Z_{n_1}$.

The ensemble of $n_2$ unitary matrices is defined by
the measure $d\nu(U_1,\dots,U_{n_2})$ on the space of
$U_1,\dots,U_{n_2}\in \mathbb{U}_N$ equal to the product of the Haar measures
$d_*U_1\cdots d_*U_{n_2}$.

In what follows in this section, we consider the mixed ensemble, which contains $n=n_1+n_2$ matrices
$Z_1,\dots,Z_{n_1}$ and $U_1,\dots,U_{n_2}$ with measure $d\Omega_{n_1,n_2}=d\mu(Z_1,\dots,Z_{n_1})d\nu(U_1,\dots,U_{n_2})$.
Let us denote the set of these $n_1+n_2$ matrices by $X$, and the related measure by $dX$. The expectation value of any function $f$ of the entries $\{X_{ij},\,1\le i,j \le N \}$
in this set of matrices is defined as
$$
\langle f \rangle = \int_{\Omega_{n_1,n_2}} f\,
dX
$$
where $\Omega_{n_1,n_2}=\mathbb{GL}^{\otimes n_1}\otimes \mathbb{U}_N^{\otimes n_2}$.

Let $\Gamma$ be an embedded graph drawn on the sphere $\mathbb{S}^2$ with $n$ band edges and $2n$ sides of these edges, numbered $\mathfrak{N}$ as described in the previous paragraph.
Then we consider the equipped graph: to the side of edge numbered $i$ ($i\in\mathfrak{N}$), we assign the matrix $X_i$, while to the opposite side of the same edge numbered $-i$, we assign the Hermitian conjugate matrix $X_{-i}:=X^\dag_i$.
Here, each $X_i$ is one of the matrices $Z_1,\dots,Z_{n_1},U_1,\dots,U_{n_2}$. Thus, the matrices
$X_1,\dots,X_n$ belong to our mixed ensemble.
Such ensembles were introduced in \cite{Alexandrov}.
Some common roots of these constructions can also be found in the long-standing work \cite{Witten}.
The solvable
cases for a single matrix (either complex or unitary) were studied earlier in \cite{O-2004-New}.

In \cite{NO2020}, \cite{NO-TMP}, the generalized model was introduced:
apart from these $2n$ matrices, we consider more $2n$ matrices $C_{\pm 1},\dots,C_{\pm n}$ equipping $\Gamma$. Each of this matrices is associated with the corner of the face polygon. Each corner is formed by one incoming and one outgoing arrow, and $C_i$ is associated with the number $i$ of the arrow incoming to the corner.

Consider a cycle $\mathfrak{f}_i=(j_1,j_2,\dots,j_{\ell_i})$, where $\ell_i$ is the number of edges
of the face $f_i$.
With each face cycle $(\mathfrak{f}_i)$, we associate two cycle products. The first one is
\be\label{face-cycle-product}
\mathfrak{f}_i\,\to\,{\cal F}_i\,:\ \ \ \ \
{\cal F}_i= C_{j_1}\cdots C_{j_{\ell_i}}\,.
\ee
which will call {\it face cycle product} (or, the same, face monodromy)
and the second is
\be\label{dressed-face-cycle product}
\mathfrak{f}_i\,\to\,{\cal F}_i(X)\,:\ \ \ \ \
{\cal F}_i(X)= X_{j_1}C_{j_1}\cdots X_{j_{\ell_i}}C_{j_{\ell_i}}\,.
\ee
which can be called {\it dressed face cycle product}, or, the same, {\it dressed face monodromy}, because it is the product of matrices
obtained by going around the face in the positive direction.

Similarly, consider a cycle
$\mathfrak{v}_i = (k_1,\dots,k_{\ell(i)})$, where $\ell_i$ is the valency of the vertex $i$. We define {\it vertex cycle product} or the same {\it vertex monodromy}, which is the product of matrices corresponding to the cycle around the vertex:
\be\label{vertex-cycle-product}
\mathfrak{v}_i\,\to\,{\cal V}_i\,:
{\cal V}_i= C_{k_1}\cdots C_{k_{\ell_i}}\,.
\ee
and in addition we define {\it dressed vertex cycle product} (or, the same, dressed vertex monodromy) as
\be\label{dressed-vertex-cycle-product}
\mathfrak{v}_i\,\to\,{\cal V}_i(X)\,:
{\cal V}_i(X)= X_{k_1}C_{k_1}\cdots X_{k_{\ell_i}}C_{k_{\ell_i}}\,.
\ee

We have wonderful dual relations
\be\label{I}
\int_{\Omega_{n_1,n_2}} \prod_{a=1}^F s_{\lambda^a}\left({\cal F}_a(X)\right) dX=\delta_\lambda
N^{-n_1d}\left(s_\lambda(\pb_\infty)\right)^{-n_1}\left(s_\lambda({I}_N)\right)^{-n_2}
\prod_{a=1}^V s_{\lambda^a}\left({\cal V}_a\right),
\ee
\be\label{II}
\int_{\Omega_{n_1,n_2}} \prod_{a=1}^V s_{\lambda^a}\left({\cal V}_a(X)\right) dX=
\delta_\lambda  N^{-n_1 d}
\left(s_\lambda(\pb_\infty)\right)^{-n_1}\left(s_\lambda({I}_N)\right)^{-n_2}
\prod_{a=1}^F s_{\lambda^a}\left({\cal F}_a\right)\,,
\ee
where $\delta_\lambda$ is equal to 1 in the case of $\lambda^1=\cdots = \lambda^F:=\lambda$ and is equal to 0 otherwise.
Here
\be
s_\lambda(\pb_\infty)=\frac{d_\lambda}{d!},\quad d_\lambda=
\frac{\prod_{i<j\le N}(\lambda_i-\lambda_j-i+j)}{\prod_{i=1}^N(\lambda_i-i+N)!},\quad
d=|\lambda|
\ee
($d_\lambda$ is the dimension of the representation $\lambda$ of the symmetric group $S_d$)
and $s_\lambda({I}_N)$ is the dimension of the representation $\lambda$ of the linear group $\mathbb{GL}_N$,
\be
s_\lambda({I}_N)=\frac{s_\lambda(\pb_\infty)}{(N)_\lambda},\quad
(N)_\lambda=\sum_{(i,j)\in\lambda}(N-i+j)=\prod_i \frac{\Gamma(N+\lambda_i-i)}{\Gamma(N-i)}\,.
\ee
Formulas (\ref{I}) and (\ref{II})
can be related, respectively, to (\ref{graph-combinatorial}) and
(\ref{graph-combinatorial*}).

\paragraph{UC mating of matrix models associated with different graphs.}

Having two different orientable surfaces and two embedded graphs, one can mate any
selected pair of the Schur functions associated with different graphs using any of
$[f \ast g]_i$, $i=1,2,3$. To apply $[f \ast g]_i$, the matrices should have the block form (\ref{block}).

The integral (\ref{J}) is the mating of two matrix models, each of them being associated with the graph with two faces, one edge and one vertex drawn on the sphere:
\be
J(X_1,X_2,\pb)=[ Z_1(X_1|\pb^1) \ast Z_2(X_2|\pb^2) ]_3,
\ee
where
\be
Z_a(X_a|\pb^a)=\int_{\mathbb{U}_N}
d_*V_a \,
 \exp \sum_{m>0}\frac 1m p^{(a)}\tr (V_aC_a)
 \exp \sum_{m>0}\frac 1m \tilde p^{(a)}\tr (V^\dag_aC_{-a})=
 \sum_{\lambda\atop\ell(\lambda)\le N} \frac{s_\lambda(\pb^a)s_\lambda(\tilde\pb^a)}{s_\lambda(I_N)}s_\lambda(C_aC_{-a}),
\ee
where the face monodromies are ${\cal F}_{1} =V_1C_1 $ (say, the Northern Hemisphere where
$V_1\in \mathbb{U}_N$ is assigned to the north side of the equator, and $C_1$ is assigned to
the northern arc of the vertex at the equator) and the monodromy of the Southern Hemisphere
is ${\cal F}_{-1} =V^\dag_1C_{-1} $, where $V^\dag_1$ is assigned to the southern side
of the equator, and $C_{-1}$ is assigned to the southern arc of the vertex. The vertex monodromy
is the product of matrices assigned to the arcs of the vertex, and is equal to
$C_1C_{-1}$.

The similar
monodromies has two faces of the second sphere: ${\cal F}_{2} =V_1C_1 $ and ${\cal F}_{-2} =V^\dag_2C_{-2} $. And the vertex monodromy is $C_2C_{-2}$. This monodromy enters the answer
for the integral (see the right hand side of (\ref{I})).

To get the UC tau function, we glue the monodromies of the vertices:
$$ [s_\lambda(C_1C_{-1})\ast s_\lambda(C_2C_{-2})]_3$$ according to Lemmas \ref{lemma2}
and \ref{lemma3} and taking into account (\ref{CaC_a}).

The model given by (\ref{J2}) is the result of mating of two two-component KP tau functions
\be
Z_a = \int_{\mathbb{U}_N} e^{\sum_{m>0} \frac 1m p^{(a)}_m
\tr \left((V_aC_aV_a^\dag C_{-a})^m\right)} d_*V_a,\quad a=1,2.
\ee
Each model is related to the graph with one edge, two vertices and one face, which is
the segment drawn on $\mathbb{S}^2$.

The first graph yields a single face monodromy, which is $V_1C_1V_1^\dag C_{-1}$,
where passing around the segment (which is the boundary of the face) one consequently meets
$V_1$ (one side of the edge), then $C_1$ assigned to the vertex, then $V_1^\dag$ assigned
to the opposite side of the edge, then $C_{-1}$ assigned to the second vertex.
The vertex monodromies are $C_1$ and $C_{-1}$, they enter the right hand side of (\ref{I}).

To get the UC integral (\ref{J2}), we mate the vertices with monodromies $C_{-1}$ and $C_{-2}$
using the block form (\ref{Ca}) and Lemmas \ref{lemma1} and \ref{lemma2}.

\section{Discussion}
We propose a way to construct UC-tau functions as the partition function of
matrix models. We used integrals over unitary and over complex matrices.
In fact, we have a huge family of matrix models with the same partition function. We have limited ourselves to
presenting the general construction and simple examples. A more detailed and complete study will be published separately.

\section*{Acknowledgements}

Our work is partly funded within the state assignment of NRC Kurchatov institute. The work of A. Mironov
is also partly supported by the grant of the Foundation for the Advancement of Theoretical Physics and Mathematics ``BASIS". Chuanzhong Li is supported by the National Natural Science Foundation of China under Grant No. 12071237 and Natural Science Research Project of Universities in Anhui Province (grant number 2024AH040202).
The work of A.O. is an output of a research project implemented as part of the Basic Research Program at the National Research University Higher School of Economics (HSE University).



\appendix

\section{Partitions. The Schur polynomials \cite{Mac} \label{Partitions}}

We recall that a nonincreasing set of nonnegative integers $\lambda_1\ge\cdots \ge \lambda_{k}\ge 0$,
we call partition $\lambda=(\lambda_1,\dots,\lambda_{l})$, and $\lambda_i$ are called parts of $\lambda$.
The sum of parts is called the weight $|\lambda|$ of $\lambda$. The number of nonzero parts of $\lambda$
is called the length of $\lambda$, it will be denoted $\ell(\lambda)$. See \cite{Mac} for details.
Partitions will be denoted by Greek letters: $\lambda,\mu,\dots$. The set of all partitions is denoted by
$\Pa$.

To define the Schur function $s_\lambda$, $\lambda\in\Pa$ at the first step we introduce the
set of elementary Schur functions $s_{(m)}$ by
\be\label{CLit}
e^{\sum_{m>0}\frac 1m p_m x^m}=\sum_{m\ge 0} x^m s_{(m)}(\pb)
\ee
where the variables $\pb=(p_1,p_2,p_3,\dots)$ are called power sum variables. For
$\lambda=(\lambda_1,\lambda_2,\dots)\in\Pa$
we define
\be
s_\lambda(\pb)=\det\left[s_{(\lambda_i-i+j)}(\pb)  \right]_{i,j>0}
\ee
Let us introduce notations $\pb_\infty:=(1,0,0,\dots)$ and $\pb(a):=(a,a,a,\dots)$.
We have (see Ch. I in \cite{Mac})
\be\label{s(p(a))}
\frac{s_\lambda(\pb(a))}{s_\lambda(\pb_\infty)}=(a)_\lambda:=\prod_{(i,j)\in\lambda}(a-j+i)
\ee
The product ranges all nodes $(i,j)$ of the Young diagram $\lambda$ \cite{Mac}.
The product on the right-hand side is called the content product, since the number $j-i$
is called the content of the node with coordinates $(i,j)$.

The Schur functions play the role of the characters of the irreducable reperesentation
labeled by $\lambda$ of the linear group.

The following relation is known as the character map relation
\be\label{charmap}
s_\lambda(X)=s_\lambda(\pb_\infty)\sum_{\mu\atop |\mu|=|\lambda|}\varphi_\lambda(\mu)
\prod_{i=1}^{\ell(\mu)}
\tr\left( X^{\mu_i} \right),\quad \pb_\infty:=(1,0,0,\dots),
\ee
where $\varphi_\lambda(\mu)$ are rational numbers which are expressed in terms of
the characters of the symmetric group $S_d,\,d=|\lambda|$, see \cite{Mac}. It follows from
(\ref{charmap}) that $s_\lambda(X)$ is a polynomial of degree $d=|\mu|=|\lambda|$ in entries
of $X$. Therefore, $s_\lambda(X)$ is a polynomial of each entry of matrix $X$ and one can integrate it over entries, as it is written in Section \ref{Ensembles}.

\section{Measures of matrix ensembles \label{Ensembles}}

\paragraph{Unitary matrices.}
The Haar measure on $\mathbb{U}_N$ is $d_*U=\prod_{i,j} \left(dU U^{-1}\right)_{ij}$ and is also written as
\be\label{Haar}
d_*U=\frac{d\Omega}{(2\pi)^N}\prod_{ 1\le i<k\le N} |e^{\sqrt{-1}\theta_i}-e^{\sqrt{-1}\theta_k}|^2 \prod_{i=1}^N d\theta_i,\quad
-\pi < \theta_1 < \cdots <\theta_N \le \pi\,.
\ee
where $e^{\sqrt{-1}\theta}$ are eigenvalues of $U\in\mathbb{U}_N$, and $d\Omega$ is the measure of the angular part (which is inessential in the case of dealing only with averages of invariant quantities), see \cite{Mehta}.

\paragraph{Ginibre ensemble of complex matrices}

\be\label{GUEmeasure'}
d\mu(Z)= Ce^{-N\Tr ZZ^\dag}\prod_{i,j\le N} d^2 Z_{i,j}\,.
\ee
where $C$ is chosen in a way $\int d\mu(Z)=1$.
The measure of ensemble of many complex matrices $Z_1,\dots,Z_n$ is the product $\prod_i^n d\mu(Z_i)$.

\section{Bilinear equations for UC hierarchy}

The pair of bilinear equations for the UC hierarchy is
\be
\sum_{k-l+m=-1} s_{k}(-2u)s_{l}(\tilde D_p)s_{m}(\tilde D_{p^*})
 \cdot\tau^{UC}(\pb+u,\pb^*+v)\tau^{UC}(\pb-u,\pb^*-v)=0,
\ee
\be
\sum_{k-l+m=-1} s_{k}(-2v)s_{l}(\tilde D_{p^*})s_{m}(\tilde D_{p})
 \cdot\tau^{UC}(\pb+u,\pb^*+v)\tau^{UC}(\pb-u,\pb^*-v)=0
\ee
where $u=(u_1,u_2,\dots)$ are arbitrary parameters, and $\tilde D_{p}=(D_1,D_2,\dots)$ are Hirota derivatives \cite{JM}.

For instance, the term $u^0v^0$ gives
$$
\sum_{m\ge 0} s_{m+1}(\tilde D_p)s_{m}(\tilde D_{p^*})\tau^{UC}(\pb,\pb^*)\tau^{UC}(\pb,\pb^*) =0
$$

If $\tau^{UC}(\pb,\pb^*)$ does not depend on $\pb^*$, one reproduces the KP Hirota bilinear equation
$$
\sum_{k\ge 0} s_{k}(-2u)s_{k+1}(\tilde D_p)
 \tau^{UC}(\pb+u)\tau^{UC}(\pb-u)=0\,.
$$

\section{Examples of graphs for matrix model \label{graphs-for-MM}}

For illustration purposes we use figures from \cite{AOV}.

  \setlength{\myStandardFigureWidth}{\linewidth}
\setlength{\subSubFigPenalty}{5mm}
\begin{arrangedFigure}{1}{3}{figure2}{ Graphs with matrices at corners and on edges }
\subFig[A fragment of a graph with random matrices on the edges and source matrices in corners ]{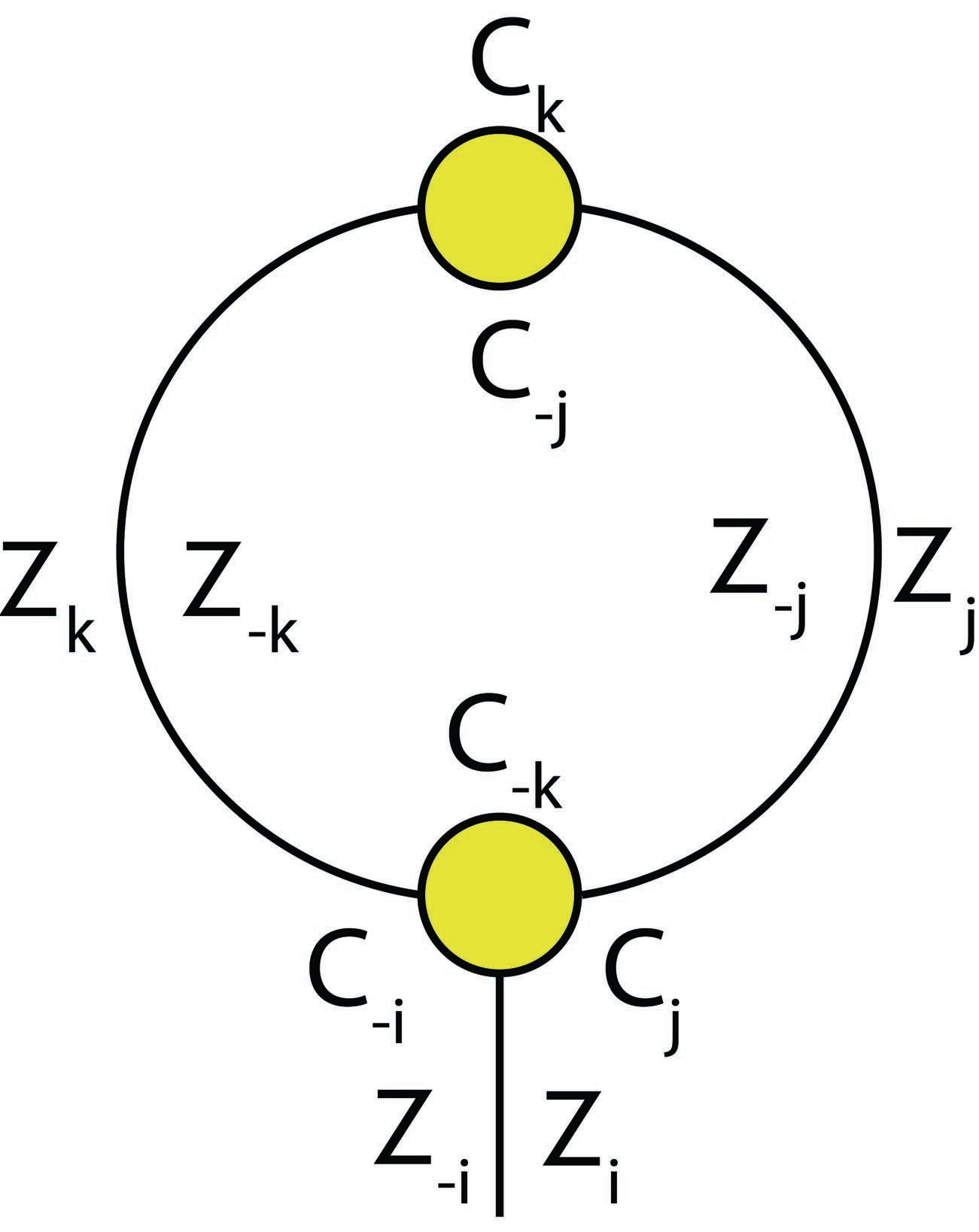}%
\subFig[1 edge (1-matrix model)]{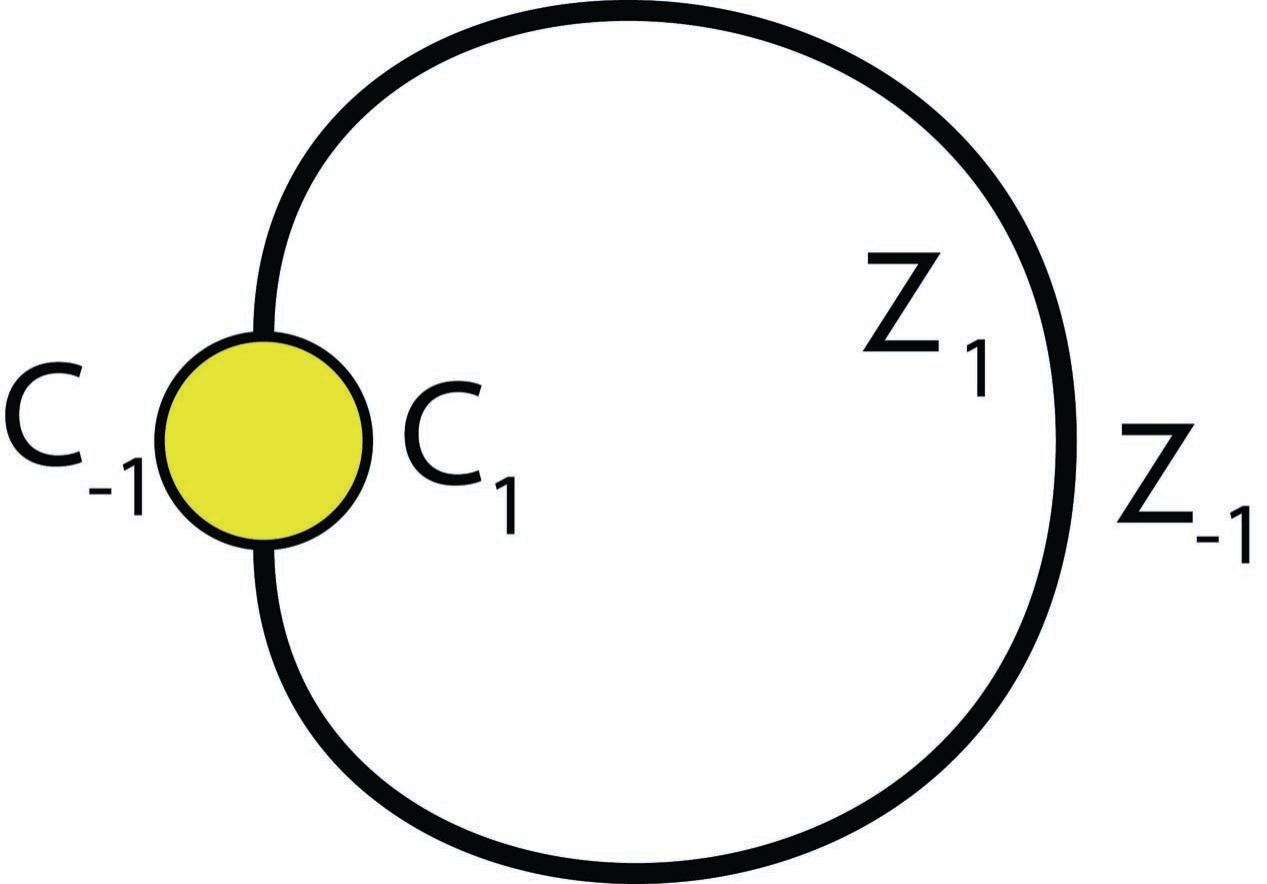}%
\subFig[1 edge (1-matrix model)]{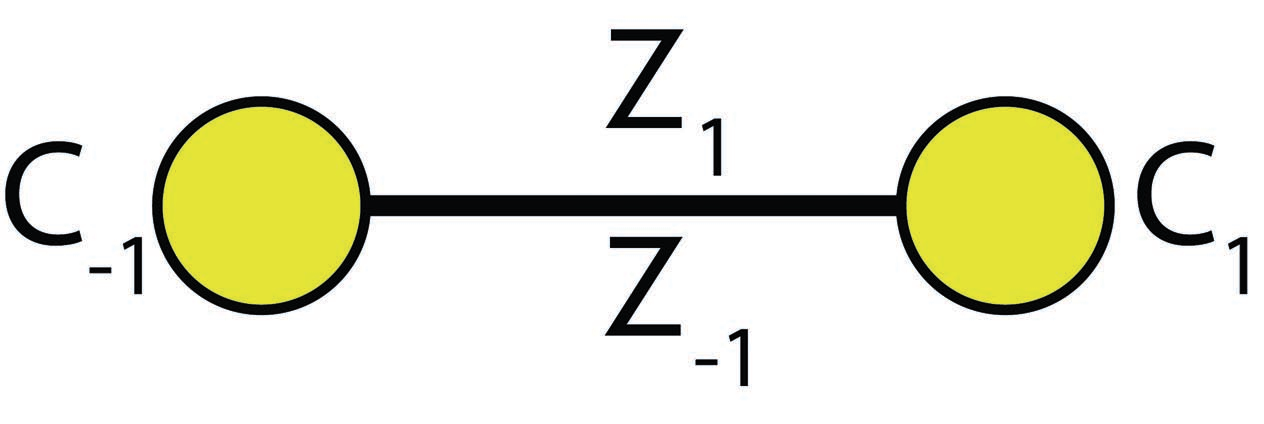}%
\end{arrangedFigure}

Figure 1a shows a fragment of a graph, which shows three edges with side numbers $(i,-i),(j,-j)$ and $(k,-k)$, as well as two vertices highlighted in yellow. Each side is assigned a random matrix with the same number: $(Z_i,Z_{-i})$ – to the sides with edge number $|i|$, $(Z_j,Z_{-j})$ – to the sides of the edge numbered $|j|$, and $(Z_k,Z_{-k})$ – to the sides with edge number $|k|$. It is assumed that $Z_{-a}=Z_a^\dag$, where $a=i,j,k$.
 If you go around a face in the positive direction, then each corner of the face is assigned the number of the previous side of the edge. In the corners, corner matrices (the same: source matrices of the model) are placed,
with the same number as the corner number. In our case, these are matrices $C_{-i}$,
following the side $-i$, then matrix $C_k$, following the side $k$, then matrix $C_j$, following the side $j$.
 We also have the corner matrix $C_{-k}$ following the side $-k$, and the corner matrix $C_{-j}$ following the side $-j$. The vertex cycles are
$\mathfrak{v}_1=(k,-j)$ and $\mathfrak{v}_2=(j,-i,-k)$, and the corresponding vertex monodromies are
$\mathfrak{V}_1=C_k,C_{-j}$ and $\mathfrak{V}_2=C_j,C_{-i},C_{-k}$ respectively.
It is not possible to write relation (\ref{I}) because it applies to the whole graph, not to a fragment. It is possible to write the cycle of only one face:
$\mathfrak{f}=(-j,-k)$, since the other faces are not represented. The monodromy of this face is $\mathfrak{F}=C_{-j}C_{-k}$, and the corresponding dressed monodromy (dressed by random matrices)  of the face is obtained from $C_a\to Z_aC_a$ and is equal to
$\mathfrak{F}(Z)=Z_{-j}C_{-j}Z_{-k}C_{-k}$

The graph in Fig. 1b gives us $\mathfrak{v}=(-1,1)$, and the corresponding vertex monodromy is $\mathfrak{V}=C_1C_{-1}$. There are two face cycles $(\mathfrak{f}_1)=(1)$ and $(\mathfrak{f}_2)=(-1)$.
The corresponding face monodromies are $\mathfrak{F}_1=C_1$ and $\mathfrak{F}_2=C_{-1}$, and the dressed
monodromies are $\mathfrak{F}_1=Z_1C_1$ and $\mathfrak{F}_2=Z_{-1}C_{-1}$.
The involution $\sigma\in S_2$ is the cycle $(1,-1)$.
We have an obvious relation (\ref{graph-combinatorial}): $\sigma \circ (\mathfrak{f}_1)(\mathfrak{f}_2)=(-1,1)$,
which desribes the "join action" which unifies the cycles $(\mathfrak{f}_1)$ and $(\mathfrak{f}_2)$. In this case, relation (\ref{I})
takes the form (\ref{integral1}).

Let us note that
the graph on Fig 1c is dual to the graph on Fig 1b. The relation (\ref{II}) reads as (\ref{lemma1a}). It is related to the ``cut action'': $(-1,1)\circ (-1,1)=(1)(-1)$.

There are three different embedded graphs with two edges and one vertex. We present the following two:
  \setlength{\myStandardFigureWidth}{\linewidth}
\setlength{\subSubFigPenalty}{5mm}
\begin{arrangedFigure}{1}{3}{figure5}{Graphs with 2 edges (two matrix models)}
\subFig[]{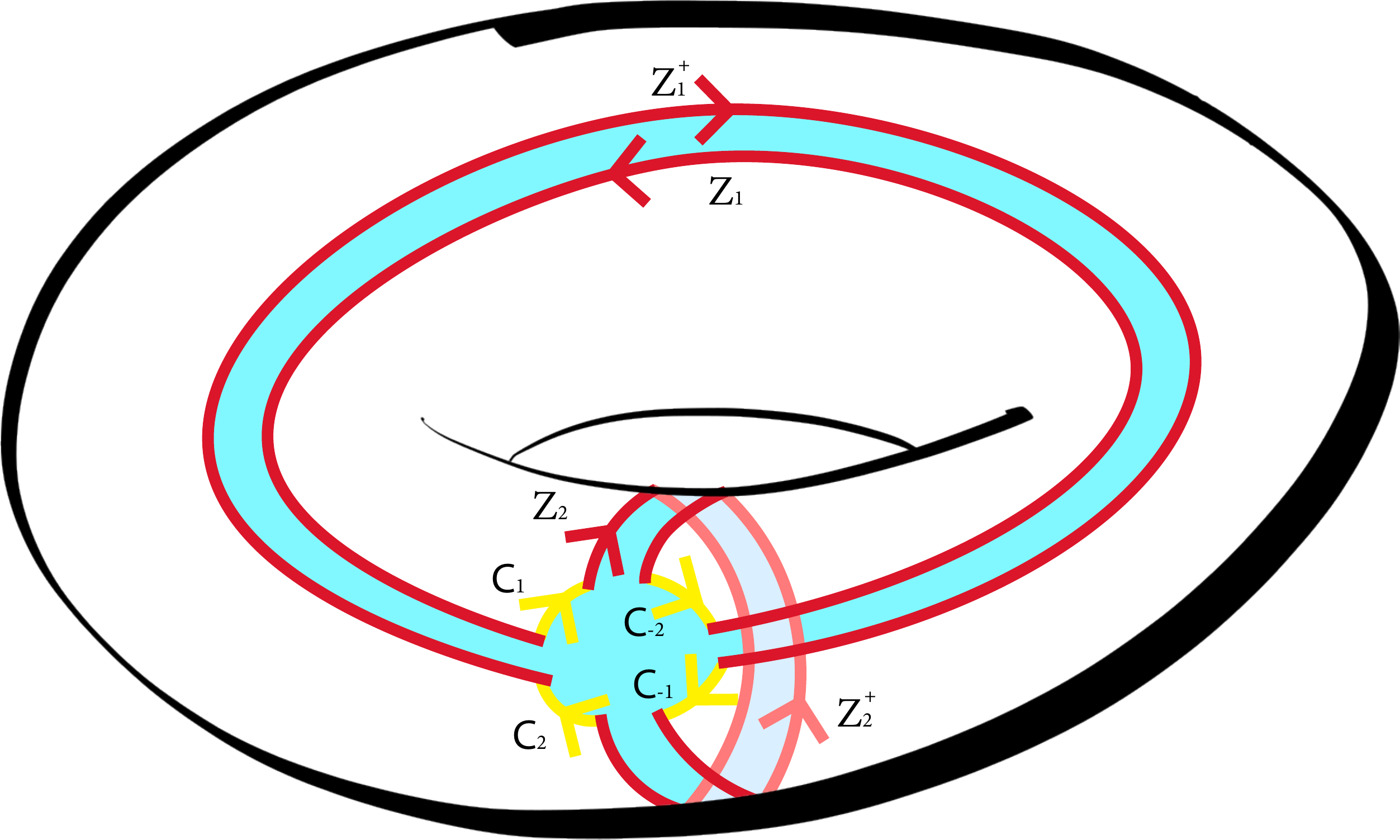}%
\subFig[]{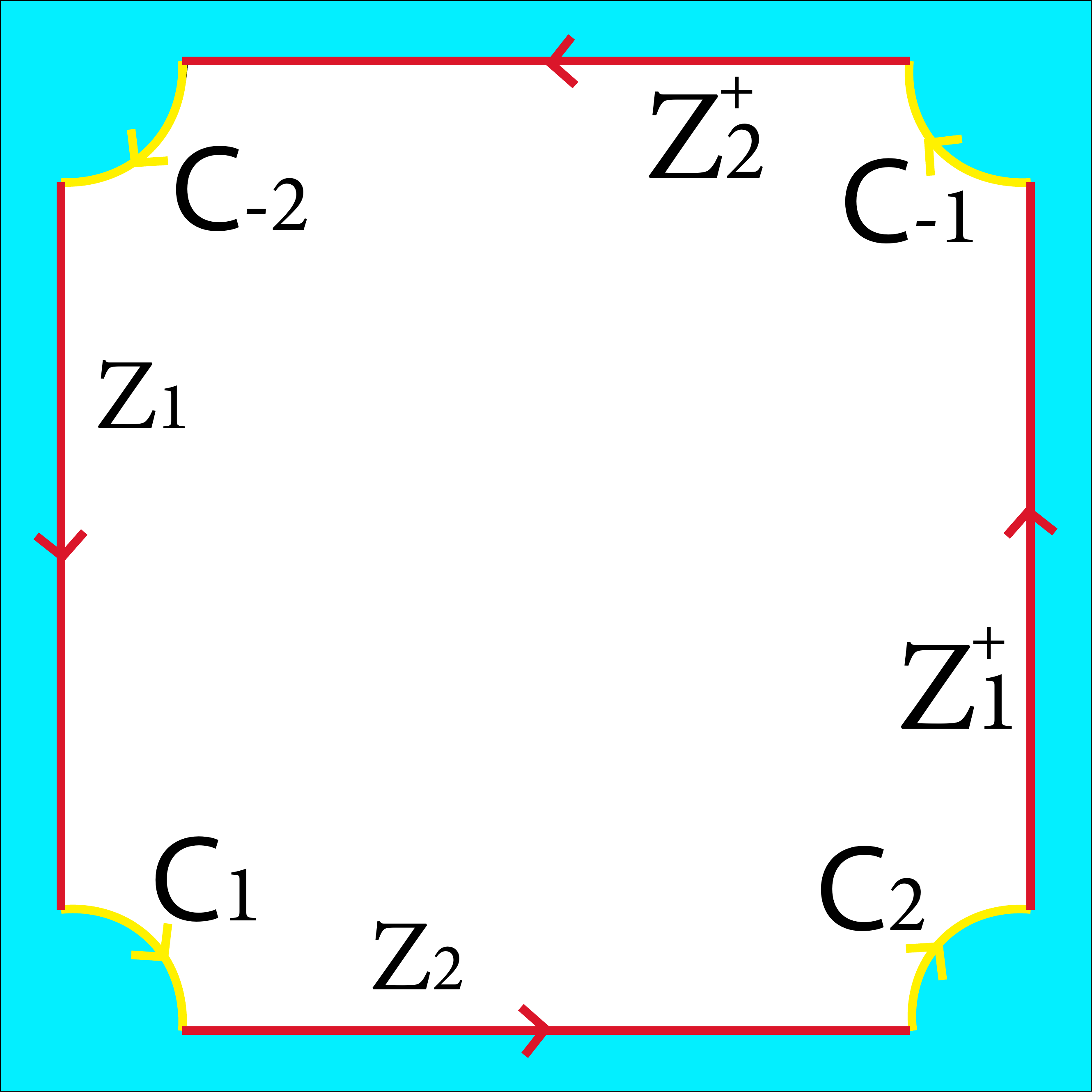}%
\subFig[]{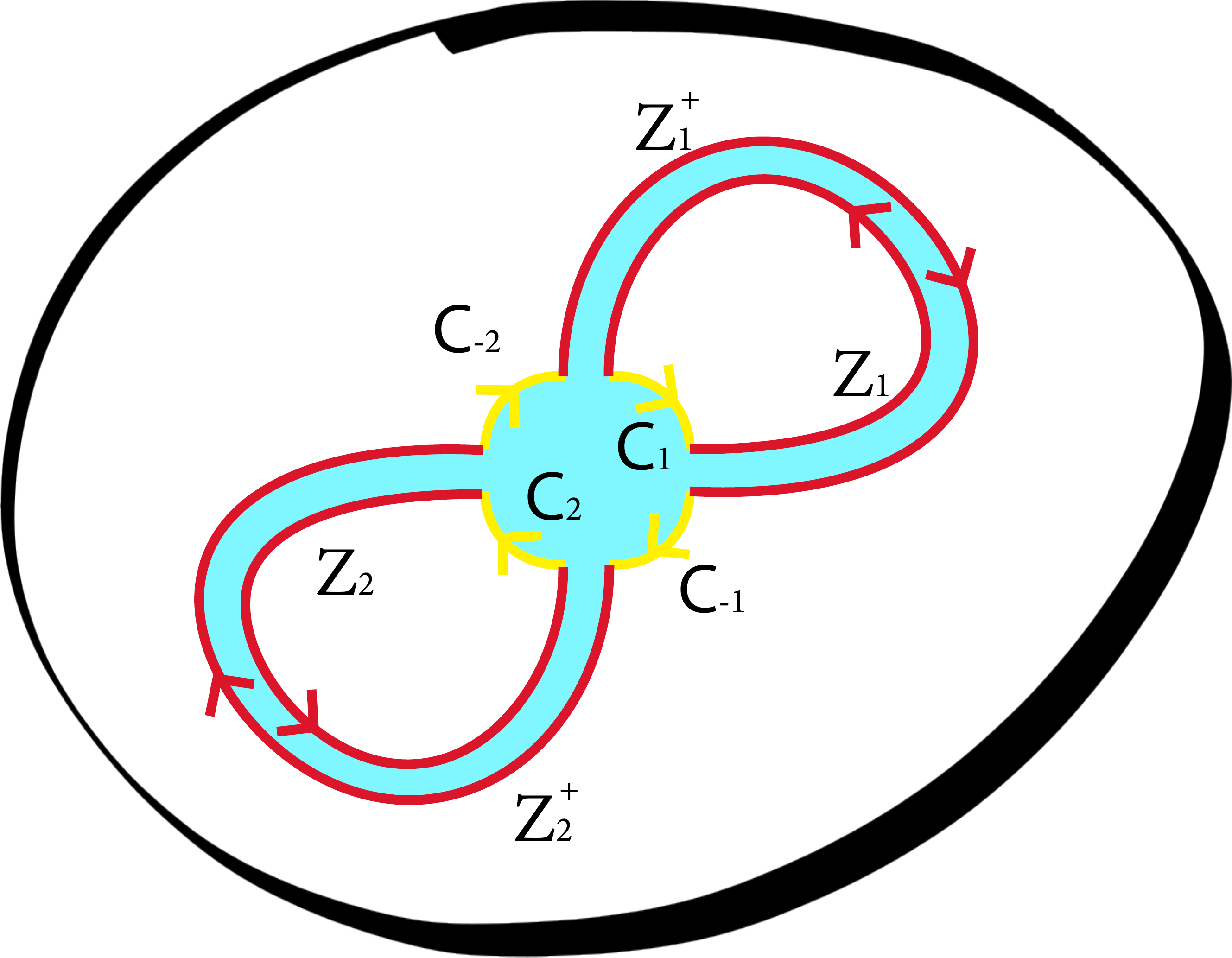}%
\end{arrangedFigure}

The graph in Fig. 2a is depicted on a torus. Its only face is the fundamental domain depicted in Fig. 2b.
The face cycle is $\mathfrak{f}=(1,2,-1,-2)$, the face monodromy is $(\mathfrak{F})=C_1C_2C_{-1}C_{-2}$, the dressed face monodromy is $\mathfrak{F}(Z)=Z_1C_1Z_2C_2Z_{-1}C_{-1}Z_{-2}C_{-2}$.
The vertex cycle is $\mathfrak{v}=(1,-2,-1,2)$.
Relation (\ref{I}) in the case of $Z\in GL_N$ has the form
$$
 \int s_\lambda\left(\mathfrak{F}(Z)\right)d\mu(Z)=N^{-2|\lambda|}s_\lambda(\pb_\infty)^{-2}s_\lambda(\mathfrak{V}),\quad
 \mathfrak{V}= C_{1} C_{-2} C_{-1} C_{2}
$$
The matrix model associated with this graph is
\be
\int e^{\sum_{m>0} N\frac{p_m}{m} \Tr\left({\mathfrak{F}(Z)}\right)^m  }d\mu(Z)=
\sum_{\lambda\atop\ell(\lambda)\le N} N^{-2|\lambda|}\frac{s_\lambda(N\pb)s_\lambda(C_{1} C_{-2} C_{-1} C_{2})}{s_\lambda(\pb_\infty)^{2}}
\ee
The right-hand side is not a tau function.

One can check that the graph in Fig 2c yields us the following matrix model
\be\label{ears}
\int \prod_{a=1}^3 e^{\sum_{m>0} N\frac{p^{(a)}_m}{m} \Tr\left({\mathfrak{F}_a(Z)}\right)^m  }d\mu(Z)=
\sum_{\lambda\atop\ell(\lambda)\le N} N^{-2|\lambda|}
\frac{s_\lambda(\mathfrak{V})\prod_{a=1}^3s_\lambda(N\pb^{(a)})}{s_\lambda(\pb_\infty)^{2}}
\ee
where $ \mathfrak{V}=C_1C_{-1}C_2C_{-2}$ and where $\mathfrak{F}_1(Z)=Z_1C_1, \,\mathfrak{F}_2(Z)=Z_2C_2,\,
\mathfrak{F}_2(Z)=Z_1^\dag C_{-1}Z_2^\dag C_{-2}$.
Under the constraints discussed in sections \ref{Ex1} and \ref{Ex2}, the right-hand side (\ref{ears}) can be equated to the tau function of the KP and TL hierarchies.

\end{document}